\documentclass[12pt]{iopart}
\usepackage{amssymb}
\usepackage{bm}
\usepackage[dvips]{graphicx}
\newcommand{\be}{\begin{equation}}
\newcommand{\en}{\end{equation}}
\newcommand{\prl}{Phys. Rev. Lett.}

\newcommand{\prb}{Phys. Rev. B}
\newcommand{\avg}[1]{\left< #1 \right>}
\newcommand{\abs}[1]{\left| #1 \right|}
\newcommand{\sr}{s_{R}}
\newcommand{\sa}{s_{A}}
\newcommand{\phir}{\phi_{R}}
\newcommand{\phia}{\phi_{A}}
\newcommand{\chir}{\chi_{R}}
\newcommand{\chia}{\chi_{A}}
\newcommand{\diag}{{\rm diag}}
\renewcommand{\Re}{{\rm Re}}
\newcommand{\tv}{\tilde{v}}
\newcommand{\hb}{\hat{B}}
\newcommand{\odet}{\overline{\det}}
\newcommand{\ket}[1]{\left| #1 \right>} 
\newcommand{\bra}[1]{\left< #1 \right|} 
\newcommand{\te}{\tilde{\epsilon}}
\newcommand{\erf}{{\rm erf}}

\begin{document}
\title{Anderson localization on a simplex}
\author{A. Ossipov}
\address{School of Mathematical Sciences, University of Nottingham, Nottingham NG7 2RD, United Kingdom} 

\begin{abstract}
We derive a field-theoretical representation for the moments of the eigenstates in the generalized  Anderson model. The representation is exact and can be used for the Anderson model with generic non-random hopping elements in any dimensions.  
We apply this method to the simplex model, for which the hopping amplitude between any two lattice sites is the same, 
and find that the eigenstates are localized at any strength of disorder. Our analytical predictions are in excellent agreement with the results of numerical simulations.
\end{abstract}

\maketitle

\section{Introduction}

The phenomenon of  Anderson localization remains one of the most important and actively studied problems in the field of disordered quantum systems. Despite a great progress in understanding of this phenomenon achieved in the last fifty years (see \cite{Abrahams} and references therein), the number of available models affordable for non-perturbative analytical treatment is very limited. Among them are the original one-dimensional Anderson model \cite{And58} and its continuous version \cite{Ber73,AR78, Mel80}, quasi-one-dimensional models and their random matrix counterparts \cite{EL83,FM91,Bee97}, as well as models on a tree-like graphs \cite{AAT73,KS80,Zir86,Ef87,Ver88,MF91}. A common feature of all such models is that they can be treated recursively. In the discrete case that means that it is possible to establish a relatively simple relation between properties of a system of size $N$ and of size $N+1$. A necessary condition for existence of such recursion relations is the absences of loops in the configuration space. Indeed, the existence of just a single loop, like in the case of the one-dimensional Anderson model with periodic boundary conditions, makes the recursive method inapplicable. 

In this paper, we study the Anderson localization in a system containing a maximum number of loops for a given number of sites. This is the Anderson model on a $d$-simplex, which is a generalization of the notion of a triangle in the two-dimensional space to the $d$-dimensional space. The Hamiltonian of the simplex model is similar to the standard Anderson model, and it consists of the hopping term and the on-site random potential:
\be\label{ham}
H=T+V, \quad T_{ij}=\frac{1}{N},\quad V_{ij}=v_i\delta_{ij},\quad i,j=1,\dots,N,
\en
where $N=d+1$ is the total number of sites, $\delta_{ij}$ is the Kronecker delta symbol and $v_i$ are independent Gaussian distributed random variables with the zero mean value $\langle v_i \rangle=0$ and the variance $\langle v_i^2 \rangle=w^2$. The hopping amplitude between any two sites is equal to $1/N$ and we set $T_{ii}=1/N$ for the sake of convenience. 

In the absence of disorder, the spectrum of the perfect system can be easily found if we notice that 
\be
Tf_0=f_0,\quad f_0=(1,1,\dots,1)^{\rm{T}},
\en
so that the vector $f_0$, whose all components are equal to one, is an eigenvector of $T$ with the eigenvalue $\lambda=1$. On the other hand, any vector $f$ orthogonal to $f_0$ satisfies the equation
\be\label{deg_states}
Tf=0.
\en
There exist $N-1$ linearly independent vectors, which are orthogonal to $f_0$, and hence we conclude that the second eigenvalue of $T$, $\lambda=0$, is $(N-1)$-fold degenerate.

Thus, the simplex model belongs to the class of the Anderson tight-binding models, whose clean analogs have flat bands of highly degenerate states. It is known that in the presence of disorder, the eigenstates of such systems may become critical at weak disorder \cite{CPS10} or exhibit the Anderson transition at strong disorder \cite{GNM06}. For the simplex model one would expect that weak disorder should lift the degeneracy and facilitate the emergence of extended eigenstates, as the hopping matrix connects any two sites of the lattice. However, we show that, in contrast to this expectation, all eigenstates, which were initially degenerate, are localized for any disorder strength in the thermodynamic limit $N\to \infty$. To this end, we calculate analytically the moments of the eigenstates as an explicit function of $w$ and demonstrate that they remain finite in the limit $N\to \infty$.
 
Another aim of this work is to present a new approach for calculation of the moments of the eigenstates suitable for Anderson models with non-trivial connectivity. Representing the disorder averaged moments of the eigenstates by the supersymmetric functional integral at the first step, we show how the most of the integration variables can be integrated out leading to the functional integral containing only a single real variable associated with every lattice site. As the constructed representation is exact and valid for the Anderson model with a generic hopping matrix, it can serve as a starting point for further investigations. In particular, for the simplex model, it can be reduced just to a two-fold integral, which can be analyzed in the limit $N\to\infty$ by the standard methods.  

The paper is organized as follows. In Section \ref{sec_general_rep}, we derive a general functional integral representation for the moments of the eigenstates of the Anderson model with a generic hopping matrix. In Section \ref{sec_simplex}, we apply this representation to the simplex model and show how in this case it can be reduced to a two-fold integral. The latter is then evaluated in the limit $N\to\infty$ and compared with the results of numerical simulations in Section \ref{sec_thermo}. Finally we conclude the paper with a discussion of the origin of the localization in the simplex model and its generalizations in Section \ref{sec_concl}.

%**************************************************************************************
\section{Moments of the eigenstates}\label{sec_general_rep}

The local moments $I_q(n)$ of the eigenstates on a given lattice site $n$ are defined as follows \cite{FM_int}
\be
I_q(n)=\frac{1}{\rho(E)}\sum_{\alpha}\avg{\abs{f_{\alpha}(n)}^{2q}\delta (E-E_{\alpha})},
\en 
where $f_{\alpha}$ is a normalized eigenstate of $H$ corresponding to an eigenvalue $E_{\alpha}$, $\rho(E)$ is the averaged density of states and $\avg{\dots}$ stands for the disorder averaging. Only the states at a given energy $E$ contribute to  $I_q(n)$ due to the presence of the $\delta$-function. The factor $1/\rho$ ensures the normalization condition $(1/N)\sum_{n=1}^N I_1(n)=1$. 

The knowledge of the moments $I_q$ as a function of the system size $N$ makes it possible to distinguish between extended and localized states. Indeed, for a completely extended state $f_{\alpha}(n)\propto 1/\sqrt{N}$ and hence $I_q\propto N^{1-q}$. On the other hand, a localized state is insensitive to increasing of the system size and therefore  $I_q\to C_q$, as $N\to\infty$, where $C_q$ is an $N$-independent constant.

The moments of the eigenstates defined above can be extracted from the singular part of a product of the diagonal elements of the retarded $G^R(E)=(E+i\epsilon-H)^{-1}$ and the advanced $G^A(E)=(E-i\epsilon-H)^{-1}$ Green's functions. Indeed, if we define $K_{l.m}(n,\epsilon)$ as
\be
K_{l,m}(n,\epsilon)=\left(G^{R}_{nn}\right)^l\left(G^{A}_{nn}\right)^m, \quad l,m=1,2,\dots,
\en
then the moments $I_q(n)$ can be found by taking the limit $\epsilon\to 0$ \cite{FM_int}
\be\label{Iq_K}
\hspace*{-40pt}I_q(n)=\frac{i^{l-m}(l-1)!(m-1)!}{2\pi\rho(E)(l+m-2)!}\:\lim_{\epsilon\to 0}\:(2\epsilon)^{l+m-1}\avg{K_{l,m}(n,\epsilon)},
\quad q=l+m.
\en
It is well known that the products of the Green's functions can be conveniently represented by the supersymmetric functional integrals \cite{Efetov}. In the case of a discrete lattice of size $N$ we introduce $N$ supervectors associated with each lattice site:
\be
\Phi_i=\left(\begin{array}{c} \sr(i) \\ \chir(i) \\ \sa(i) \\ \chia(i)  \end{array}\right),\quad i=1,\dots,N,
\en
whose components are two complex $\sr$, $\sa$ and two Grassmann $\chir$, $\chia$ variables. The product of the diagonal elements of the Green's functions can be written now as the Gaussian integral over supervectors $\Phi_i$ \cite{FM_int,Efetov}:
\begin{eqnarray}\label{K-notav}
K_{l,m}(n,\epsilon)&=&\frac{i^{l-m}}{l!\:m!}\int\prod_{p=1}^N d\Phi_p d\Phi^{\dagger}_p\: 
(\sr^{\ast}(n)\sr(n))^l (\sa^{\ast}(n)\sa(n))^m\nonumber\\
&&\exp\left[i\sum_{p,q=1}^N (H_{pq}-E\delta_{pq})(\Phi_p,\Phi_q)-\epsilon\sum_{p=1}^N(\Phi_p,\Lambda\Phi_p)\right],
\end{eqnarray} 
where $\Phi^{\dagger}=(\sr^{\ast},\chir^{\ast},\sa^{\ast},\chia^{\ast})$, 
$d\Phi d\Phi^{\dagger}=-\frac{1}{\pi^2}d^2\sr d^2\sa d\chir d\chir^{\ast}d\chia d\chia^{\ast}$, the diagonal matrix $\Lambda=\diag(1,1,-1,-1)$ and we define the scalar product of two supervectors as
\be
(\Phi_p,\Phi_q)=\sr^{\ast}(p)\sr(q)+\chir^{\ast}(p)\chir(q)-\sa^{\ast}(p)\sa(q)+\chia^{\ast}(p)\chia(q).
\en
This representation enables us to perform averaging over the random diagonal part of the Hamiltonian (\ref{ham}) explicitly. Then the action of the functional integral (\ref{K-notav}) takes the form
\be\label{action}
\hspace*{-50pt}S[\Phi,\Phi^{\dagger}]=
\sum_{p=1}^N\left(\frac{w^2}{2}(\Phi_p,\Phi_p)^2+iE(\Phi_p,\Phi_p)+\epsilon(\Phi_p,\Lambda\Phi_p)\right)
-i\sum_{p,q=1}^N T_{pq}(\Phi_p,\Phi_q).
\en
It is convenient to introduce the generating function $Y(\Phi_n,\Phi_n^{\dagger})$, which is obtained by integrating $\exp(-S[\Phi,\Phi^{\dagger}])$ over all $\Phi_p$ except $\Phi_n$:
\be\label{gen-fun}
Y(\Phi_n,\Phi_n^{\dagger})=\int\prod_{p\neq n} d\Phi_p d\Phi^{\dagger}_p\:e^{-S[\Phi,\Phi^{\dagger}]}.
\en
The lower and the upper limits for $p$ is not written explicitly in order to lighten the notation. The averaged  products of the Green's functions can be now calculated by integrating the generating function
\be\label{K-avg}
\hspace*{-50pt}\avg{K_{l,m}(n,\epsilon)}=\frac{i^{l-m}}{l!\:m!}\int d\Phi_n d\Phi^{\dagger}_n\:
(\sr^{\ast}(n)\sr(n))^l (\sa^{\ast}(n)\sa(n))^m Y(\Phi_n,\Phi_n^{\dagger}).
\en
We would like to stress that Eqs.(\ref{action},\ref{gen-fun},\ref{K-avg}) are valid for an arbitrary hopping matrix $T$. There are only two properties that we require from $T$ at this point: i) $T$ is real and symmetric; ii) all the diagonal elements of $T$ are zero. They allow us to rearrange the hopping term in the action (\ref{action})
\begin{eqnarray}\label{hopping}
\sum_{p,q} T_{pq}(\Phi_p,\Phi_q)&=&
\sum_{p,q}T_{pq}\Re \left(\sr^{\ast}(p)\sr(q)-\sa^{\ast}(p)\sa(q)\right)+\nonumber\\
&& \sum_{p,q}T_{pq}\left(\chir^{\ast}(p)\chir(q)+\chia^{\ast}(p)\chia(q)\right).
\end{eqnarray}
The supersymmetric representation of the generating function constructed above is exact, but rather complicated, as it involves $8$ integration variables ($4$ real and $4$ Grassmann) associated with each lattice site. Our aim is to demonstrate that in the limit $\epsilon\to 0$ one can integrate out $7$ out of $8$ variables exactly. This can be done by changing the variables, which makes it possible to take the limit $\epsilon\to 0$ explicitly in the action of the functional integral. A new functional integral representation, obtained in such a way, is still exact, but requires an integration over a single real variable on each lattice site. 

The first step in realizing this program is to introduce the modulus and the phase of the complex variables 
$\sr=\abs{\sr}e^{i\phir}$ and $\sa=\abs{\sa}e^{i\phia}$. Then the non-Grassmann part of the hopping term (\ref{hopping})
can be written as
\begin{eqnarray}\label{cos}
\hspace*{-60pt}\sum_{p, q}T_{pq}\Re \left(\sr^{\ast}(p)\sr(q)-\sa^{\ast}(p)\sa(q)\right)&=&
\sum_{p,q}T_{pq}\left(\cos(\phir(p)-\phir(q))\abs{\sr(p)}\abs{\sr(q)}\right.-\nonumber\\
&&\left.\cos(\phia(p)-\phia(q))\abs{\sa(p)}\abs{\sa(q)}\right).
\end{eqnarray}
Since only the differences between phases appear in the action, we may get rid of the phases associated with the $n$th site by shifting the variables $\phir(p)\to \phir(p)+\phir(n)$ and $\phia(p)\to \phia(p)+\phia(n)$ for all $p\neq n$. This shift does not change the form of the action, but it leads to an additional constraint $\phir(n)=\phia(n)=0$.

The $\epsilon$-dependence of the action (\ref{action}) suggests that in the limit $\epsilon\to 0$ the main contribution to the integrals comes from the regions $\abs{\sr},\:\abs{\sa}\sim 1/\sqrt{\epsilon}$. Therefore it is convenient to define two new variables
\be
s_p=\epsilon(\abs{\sr(p)}^2+\abs{\sa(p)}^2),\quad v_p=\abs{\sr(p)}^2-\abs{\sa(p)}^2,
\en
so that the original variables can be expanded in a power series in $\epsilon$:
\be
\hspace*{-40pt}\abs{\sr(p)}=\sqrt{\frac{s_p}{2\epsilon}}\left(1+\frac{\epsilon v_p}{2s_p}+O(\epsilon^2)\right),\quad 
\abs{\sa(p)}=\sqrt{\frac{s_p}{2\epsilon}}\left(1-\frac{\epsilon v_p}{2s_p}+O(\epsilon^2)\right),
\en
where the terms of the order of $\epsilon^2$ and higher play no role in the limit $\epsilon\to 0$. Substituting these two expressions into Eq.(\ref{cos}), one can notice the presence of the large factor $1/\epsilon$ in the action, which enable us to integrate over the phases using the stationary phase approximation. The details of the calculation are presented in \ref{app_phase} and the result reads
\be\label{J}
J=(2\pi\epsilon)^{(N-1)}\sum_{\{\sigma_p\}}e^{i\sum_{p,q}T_{pq}\sigma_p \sigma_q v_p\sqrt{\frac{s_q}{s_p}}}
\frac{1}{\det B}\:\prod_{p\neq n}\frac 1s_p,
\en
which is the result of the integration of the  non-Grassmann part of the hopping term (\ref{cos}) over all phase variables. The matrix $B$ appearing in the above equation is defined as
\be\label{B_matr}
\hspace*{-40pt}B_{pq}=
-T_{pq}+\delta_{pq}\sum_r T_{pr}\sigma_p\sigma_r\sqrt{\frac{s_r}{s_p}},
\quad p,q=1,\dots,N;\;p,q\neq n.
\en 
The discrete variables $\sigma_p=\pm 1$ describe all possible configurations of the stationary phase points. Eq.(\ref{J}) represents the leading order in $\epsilon$ result for the integral, all higher order terms give no contribution to the generating function in the limit $\epsilon\to 0$.

The next step is to integrate out the Grassmann variables. However, the appearance of the quartic terms in the diagonal part of the action (\ref{action}) prevent us from reaching this aim directly. In order to get rid of the quartic terms in the Grassmann variables we make a shift of the variables $v_p$:
\be
\tv_p=v_p+\chir^{\ast}(p)\chir(p)+\chia^{\ast}(p)\chia(p).
\en
This allows us to write the diagonal part of the action (\ref{action}) as
\be\label{S_diag}
S_{{\rm diag}}=\sum_{p}\left(\frac{w^2 }{2}\tv_p^2+iE\tv_p+s_p\right),
\en
where in the last term we retain only the contribution, which gives a non-zero result in the limit $\epsilon\to 0$.

The shift of $v_p$ produces  new Grassmann terms in Eq.(\ref{J}), which together with the Grassmann part of the hopping term 
(\ref{hopping}), yields the following integral over Grassmann variables:
\be\label{G_int}
%\hspace*{-60pt}G=\int \prod_{p\neq n}d\chir (p) d\chir^{\ast}(p)d\chia (p) d\chia^{\ast} (p) e^{i\sum_{p,q}
%\left(T_{pq}-\delta_{pq}\sum_{r}T_{pr}\sigma_p\sigma_r \sqrt{\frac{s_r}{s_p}}\right)
%\left(\chir^{\ast}(p)\chir(q)+\chia^{\ast}(p)\chia(q)\right)}
G=\int \prod_{p\neq n}[d\chi(p)]e^{i\sum_{p,q}
\left(T_{pq}-\delta_{pq}\sum_{r}T_{pr}\sigma_p\sigma_r \sqrt{\frac{s_r}{s_p}}\right)
\left(\chir^{\ast}(p)\chir(q)+\chia^{\ast}(p)\chia(q)\right)},
\en
where $[d\chi(p)]=d\chir (p)\: d\chir^{\ast}(p)\:d\chia (p)\: d\chia^{\ast} (p)$. The integral now is Gaussian, but its action contains the Grassmann variables associated with the $n$th site, which are not integrating out in $G$. One can eliminate these variables from the integrand by a change of the variables similar to the one used in the integration over the phases. Once it is done, the integral becomes the standard Gaussian one, and the result reads (see \ref{app_grass} for details)
\be\label{G}
G=(-1)^{N-1}(\det B)^2.
\en
It is remarkable that the integration over the Grassmann variables leads to the appearance of the  determinant of exactly the same matrix $B$ as the integration over the phases.

Collecting the results from Eqs.(\ref{J}), (\ref{S_diag}) and (\ref{G}) and taking into account the expression for the measure $d\Phi d\Phi^{\dagger}=-\frac{1}{8\pi^2\epsilon}ds\: dv\:d\phi_R d\phi_A d\chir d\chir^{\ast}d\chia d\chia^{\ast}$, we obtain
\be
\hspace*{-62pt}Y(s_n,\tv_n)=\sum_{\{\sigma_p\}}\prod_{p\neq n}\int_0^{\infty}\frac{ds_p}{4\pi s_p}
\int_{-\infty}^{\infty}d\tv_p \det B\; e^{-\sum_p \left(\frac{w^2}{2}\tv_p^2+iE\tv_p+s_p\right)+
i\sum_{p,q}T_{pq}\sigma_p \sigma_q \tilde{v}_p\sqrt{\frac{s_q}{s_p}}}.
\en
The variables $\tv_p$ can be easily integrated out, as all the integrals over $\tv_p$ are Gaussian. Finally, we introduce the new variables $t_p=\sigma_p\sqrt{s_p}$, such that $\sum_{\sigma_p}\int_0^{\infty}\frac{ds_p}{s_p}=2\int_{-\infty}^{\infty}\frac{dt_p}{t_p}$, then the expression for the generating function integrated over $\tv_n$ reads
\be\label{gen_fun_final}
\hspace*{-60pt}Y(t_n)=\int_{-\infty}^{\infty}d\tv_n Y(t_n,\tv_n)=\frac{\sqrt{2\pi}}{w}
\prod_{p\neq n}\int_{-\infty}^{\infty}\frac{dt_p}{\sqrt{2\pi} w t_p}\det B 
e^{-\sum_p \left(\frac{\left(\sum_qT_{pq}\frac{t_q}{t_p}-E\right)^2}{2w^2}+t_p^2\right)}.
\en
From Eqs.(\ref{Iq_K}) and (\ref{K-avg}) it follows that the moments $I_q(n)$ are related to $Y(t_n)$ as
\be\label{mom_gen_final}
I_q(n)=\frac{1}{\pi\rho(E)(q-2)!}\int_{0}^{\infty}dt_n\: t_n^{2q-3}Y(t_n).
\en
Eqs.(\ref{gen_fun_final}), (\ref{mom_gen_final}) along with the definition of $B$ in terms of $t_p$
\be\label{B_matr_t}
\hspace*{-40pt}B_{pq}=
-T_{pq}+\delta_{pq}\sum_r T_{pr}\frac{t_r}{t_p},
\quad p,q=1,\dots,N;\;p,q\neq n,
\en 
represent the main result of this section.

The constructed representation for the moments $I_q(n)$ is exact and valid for a generic hopping matrix $T_{pq}$. In particular, it can be used to study the original Anderson tight-binding model in any dimensions. The main advantage of this approach in comparison to the initial representation in terms of the supervectors (\ref{action}) or the supersymmetric non-linear $\sigma$-model \cite{Efetov} is that the action depends on a single real-valued field $t_p$. Generally, the main obstacle to further analysis is the presence of the functional determinant, which makes the action non-local. However, for some models, the functional determinant can be calculated exactly. One example of this kind is the one-dimensional Anderson model. In this case, one can show that the action of the above representation can be reduced in the continuous limit to the action of the Liouville theory, reproducing the results for $I_q(n)$ obtained by the different methods \cite{Kolokolov,OK06}. The Anderson model on a $d$-simplex, considered in the next Section, is another example, for which the derived representation turns out to be very useful. 
%**************************************************************************************

\section{Anderson model on a simplex}\label{sec_simplex}

The general approach presented in the previous Section can be applied to the simplex model. Since in the derivation of the formula for the generating function (\ref{gen_fun_final}) we used the assumption that $T_{pp}=0$, we need to take into account the non-zero diagonal elements of the Hamiltonian (\ref{ham}) by shifting the energy $E\to E-1/N$ in Eq.(\ref{gen_fun_final}).
 Once it is done, the generating function takes the following form
\be
\hspace*{-0pt}Y(t_n)=\frac{\sqrt{2\pi}}{w}
\prod_{p\neq n}\int_{-\infty}^{\infty}\frac{dt_p}{\sqrt{2\pi} w t_p}\det B 
e^{-\sum_p \left(\frac{\left(\frac{1}{N}\sum_q\frac{t_q}{t_p}-E\right)^2}{2w^2}+t_p^2\right)}.
\en
We are interested in the eigenstates, which are initially degenerate at $w=0$. They correspond to $E=0$ and this is the value of the energy we are going to consider below\footnote{The averaged density of states is exponentially small at $w\to 0$ at any other value of energy, as it follows from \ref{app_dos}. The results for any $E$, such that $\abs{E}<w$, must be qualitatively the same as for $E=0$. The fate of the single extended state at $E=1$ in the presence of disorder is an interesting problem, which is not considered in this work. Such a gapped state has a similar origin to Cooper pairs in the simplest models of superconductors \cite{CF09}.}. The functional determinant appearing in the above equation is calculated exactly 
in \ref{app_det} and given by Eq.(\ref{det_simplex}). Using this result we obtain
\begin{eqnarray}
I_q(N)&=&c_N\int_{0}^{\infty}dt_n\: t_n^{2q-2}
\prod_{p\neq n}\int_{-\infty}^{\infty}\frac{dt_p}{t_p^2}
\left(\sum_{r=1}^N t_r \right)^{N-2}
e^{-\sum_p \left(\frac{\left(\sum_q t_q \right)^2}{2N^2 w^2 t_p^2}+t_p^2\right)},\nonumber\\
c_N&=&\frac{2}{\rho(0)(q-2)!N^{N-1}(\sqrt{2\pi} w)^N}.
\end{eqnarray}
All the sites are equal in the simplex model and therefore we can skip $n$-dependence in the notation for the moments.
At the same time, we stress the dependence of $I_q$ on the total number of sites $N$.

Analyzing the action of the functional integral above, we can notice that $\sum_q t_q$ plays the role of a collective variable, which determines the behavior of the integral. In particular, the condition $\sum_q t_q=0$ gives a solution of the saddle-point equation in the limit $w\to 0$. This condition can be viewed as a functional integral counterpart of Eq.(\ref{deg_states}), which determines the degenerate eigenstates at $E=0$. For this reason it is convenient to introduce a new variable $s$ by inserting the decomposition of unity $1=\int_{-\infty}^{\infty}ds\: \delta(s-(1/w N)\sum_q t_q)$ in the functional integral:
\begin{eqnarray}
\hspace*{-50pt}I_q(N)&=&\tilde{c}_N\int_{-\infty}^{\infty}dt_n\: t_n^{2q-2}
\prod_{p\neq n}\int_{-\infty}^{\infty}\frac{dt_p}{t_p^2}\int_{-\infty}^{\infty} ds\:
\delta\left(s-\sum_q \frac{t_q}{wN}\right)
s^{N-2}
e^{-\sum_p \left(\frac{s^2}{2 t_p^2}+t_p^2\right)},\nonumber\\
\hspace*{-50pt}\tilde{c}_N&=&\frac{1}{\rho(0)(q-2)!N(\sqrt{2\pi})^N w^2}.
\end{eqnarray}
Here we extended the lower limit of the integration over $t_n$ to $-\infty$ by assuming that $N$ is even.
Now we scale the integration variables $t_p=s x_p$ and represent the delta function by the integral
\be
\delta\left(s-\sum_q \frac{t_q}{wN}\right)=\frac{wN}{2\pi\abs{s}}\int_{-\infty}^{\infty}d\theta 
e^{-i\theta(wN-\sum_q x_q)}.
\en
This allows us to perform the integration over all variables $x_p$ separately. As a result, we obtain a very compact representation for $I_q$:
\begin{eqnarray}\label{simp_mom_fin}
\hspace*{-60pt}&&I_q(N)=r_q\int_{-\infty}^{\infty}d\theta e^{-i\theta w N}\int_{-\infty}^{\infty}ds 
\abs{s}^{2q-3}f^{N-1}(s,\theta)\:g(s,\theta),\quad 
r_q=\frac{1}{(2\pi)^{\frac{3}{2}}\rho(0)w(q-2)!},\nonumber\\
\hspace*{-60pt}&&f(s,\theta)=\int_{-\infty}^{\infty}\frac{dx}{\sqrt{2\pi}}\:x^{-2}e^{-\frac{1}{2x^2}-s^2x^2+i\theta x},\quad
g(s,\theta)=\int_{-\infty}^{\infty}dx\:x^{2q-2}e^{-\frac{1}{2x^2}-s^2x^2+i\theta x}.
\end{eqnarray}
Eq.(\ref{simp_mom_fin}) is exact and it can be used to calculate $I_q$ for any finite (even) number of sites $N$. On the other hand, the explicit $N$ dependence makes it possible to analyze the integral in the limit $N\to\infty$ and this is the content of the next Section.

%**************************************************************************************
\section{Moments of the eigenstates in the thermodynamic limit $N\to \infty$}\label{sec_thermo}

In order to find an asymptotic behavior of the double integral (\ref{simp_mom_fin}) as $N\to\infty$ we notice that the function 
\be
u(\theta)=\int_{-\infty}^{\infty}ds \abs{s}^{2q-3}f^{N-1}(s,\theta)\:g(s,\theta)
\en
has a maximum at $\theta=0$. Indeed, it follows from the definitions of $f(s,\theta)$ and $g(s,\theta)$ that 
$\abs{f(s,\theta)}\le f(s,0)$ and $\abs{g(s,\theta)}\le g(s,0)$, hence
\be
\abs{u(\theta)}\le\int_{-\infty}^{\infty}ds \abs{s}^{2q-3}\abs{f(s,\theta)}^{N-1}\:\abs{g(s,\theta)}\le u(0).
\en 
It means that in the limit $N\to \infty$ the main contribution to the integral over $\theta$ in Eq.(\ref{simp_mom_fin}) originates from $\theta\to 0$. Therefore it is convenient to introduce the new integration variables $\alpha=N \theta$ and $t=Ns$ in Eq.(\ref{simp_mom_fin}):
\be\label{Iq_N}
\hspace*{-20pt}I_q(N)=\frac{r_q}{N^{2q-1}}\int_{-\infty}^{\infty}d\alpha e^{-i\alpha w }\int_{-\infty}^{\infty}dt 
\abs{t}^{2q-3}f^{N-1}\left(\frac tN,\frac{\alpha}{N}\right)\:g\left(\frac tN,\frac{\alpha}{N}\right).
\en
Now we need to find the asymptotic expressions for $f(\epsilon t, \epsilon \alpha)$ and $g(\epsilon t, \epsilon \alpha)$ as $\epsilon\to 0$. For $g$ it can be done just by changing the integration variable $x$ in Eq.(\ref{simp_mom_fin}) by $y=x/N$:
\be\label{y_int}
g\left(\frac tN,\frac{\alpha}{N}\right)=N^{2q-1}\left[\int_{-\infty}^{\infty}dy\:y^{2q-2}e^{-y^2t^2+i\alpha y}+O(N^{-2})\right].
\en
The integral over $y$ can be now explicitly calculated in terms of elementary functions
\begin{eqnarray}\label{g_exp}
&&g\left(\frac tN,\frac{\alpha}{N}\right)=N^{2q-1}\left[\abs{t}^{-2q+1}F_q\left(\frac{\alpha}{2t}\right)
+O(N^{-2})\right],\\
\label{def_F}&&F_q(z)=\sqrt{\pi}e^{-z^2}\sum_{p=0}^{q-1}2^p(-z^2)^{q-1-p}\frac{(2q-2)!}{p!(2q-2-2p)!}.
\end{eqnarray}
For $f(\epsilon t, \epsilon \alpha)$ more careful analysis is required, as the function is not analytic at $\epsilon=0$. Its asymptotic expansion at $\epsilon\to 0$ can be written as
\be
f(\epsilon t, \epsilon \alpha)=f(0,0)+\epsilon\lim_{\epsilon\to 0}f_{\epsilon}^{\prime}(\epsilon t, \epsilon \alpha)+O(\epsilon^2).
\en 
Calculating the derivative $f_{\epsilon}^{\prime}(\epsilon t, \epsilon \alpha)$ using the integral representation for $f$, which is given by Eq.(\ref{simp_mom_fin}), we obtain
\be
f(\epsilon t, \epsilon \alpha)=1-\sqrt{2}\abs{\epsilon}\abs{t}e^{-\left(\frac{\alpha}{2 t}\right)^2}
-\sqrt{\frac{\pi}{2}}\abs{\epsilon}\abs{\alpha}\erf \left(\abs{\frac{\alpha}{2t}}\right)
+O(\epsilon^2).
\en
Applying this expansion with $\epsilon=1/N$ to $f^{N-1}\left(\frac tN,\frac{\alpha}{N}\right)$ we find
\be\label{f_exp}
f^{N-1}\left(\frac tN,\frac{\alpha}{N}\right)=e^{-\sqrt{2}\abs{t}e^{-\left(\frac{\alpha}{2 t}\right)^2}
-\sqrt{\frac{\pi}{2}}\abs{\alpha}\erf \left(\abs{\frac{\alpha}{2t}}\right)}+O(N^{-1}).
\en
Now we can substitute the results of Eqs.(\ref{g_exp}) and (\ref{f_exp}) into Eq.(\ref{Iq_N}) and take the limit $N\to\infty$
\begin{eqnarray}
\hspace*{-60pt} I_q&=&\lim_{N\to \infty}I_q(N)=r_q \int_{-\infty}^{\infty}\frac{dt}{t^2} \int_{-\infty}^{\infty}d\alpha\:
F_q\left(\frac{\alpha}{2t}\right)e^{-i\alpha w-\sqrt{2}\abs{t}e^{-\left(\frac{\alpha}{2 t}\right)^2}
-\sqrt{\frac{\pi}{2}}\abs{\alpha}\erf \left(\abs{\frac{\alpha}{2t}}\right) }=\nonumber\\
\hspace*{-60pt}&&\frac{1}{\pi (q-2)!} \int_{-\infty}^{\infty}\frac{dt}{\abs{t}}\int_{-\infty}^{\infty}dz\:
F_q(z)\:e^{-2i zt w-\sqrt{2}\abs{t}e^{-z^2}
-\sqrt{2\pi}\abs{zt}\erf (\abs{z}) },
\end{eqnarray}
where in the last line we changed the variable $\alpha$ by $z=\alpha/2t$ and substituted into the expression for $r_q$ the density of states calculated in \ref{app_dos}.

Since the integral over $t$ is simpler than the integral over $z$, we would like to change the order of the integration and to integrate first over $t$. However, the integral over $t$ diverges at $t\to 0$, so it should be first regularized. One way to do it is to replace $1/\abs{t}$ term by $1/\abs{t}^{1-\delta}$, integrate over $t$ and then take the limit $\delta\to0$. This procedure yields
\be\label{Iq_final}
\hspace*{-60pt}I_q=-\frac{1}{\pi}\int_{-\infty}^{\infty}dz\:\frac{F_q(z)}{(q-2)!}
\ln\left[4z^2w^2+2\left(e^{-z^2}+\sqrt{\pi}\abs{z}\erf (\abs{z})\right)^2 \right],\quad q=2,3,\dots.
\en
Eq.(\ref{Iq_final}) and Eq.(\ref{def_F}) represent the main result of this Section. They allow us to calculate the positive integer moments $I_q$ at arbitrary value  of $w$. In particular, the fact that the moments remain finite in the limit $N\to \infty$ means that the eigenstates are localized at any disorder strength $w$. 

One can check that in limit $w\to\infty$ Eq.(\ref{Iq_final}) gives $I_q=1$ for any $q$, which is in agreement with the expectation that at very strong disorder the states are localized at single sites of the lattice. On the other hand, the fact that Eq.(\ref{Iq_final}) predicts non-trivial values for $I_q$ at $w\to 0$ is not so obvious.

Although in our derivation, we assumed that $q$ is an integer, one can easily generalize the result for $I_q$ to non-integer values of $q$. To this end, one needs to calculate the integral in Eq.(\ref{y_int}) assuming that $q$ is non-integer and to replace $(q-2)!$ in Eq.(\ref{Iq_final}) by $\Gamma(q-1)$:
\begin{eqnarray}\label{Iq_all_q}
&& \hspace*{-40pt}I_q=-\frac{1}{\pi}\int_{-\infty}^{\infty}dz\:\frac{\tilde{F}_q(z)}{\Gamma(q-1)}
\ln\left[4z^2w^2+2\left(e^{-z^2}+\sqrt{\pi}\abs{z}\erf (\abs{z})\right)^2 \right],\quad q>1,\nonumber\\
&&\hspace*{-40pt} \tilde{F}_q=\Gamma\left(q-\frac12\right)\:{}_1F_1\left(q-\frac12, \frac12, -z^2\right),
\end{eqnarray}
where ${}_1F_1(a,b,z)$ is the Kummer confluent hypergeometric function. 

\begin{figure}[t]
\begin{center}
\includegraphics[clip=true,width=\columnwidth]{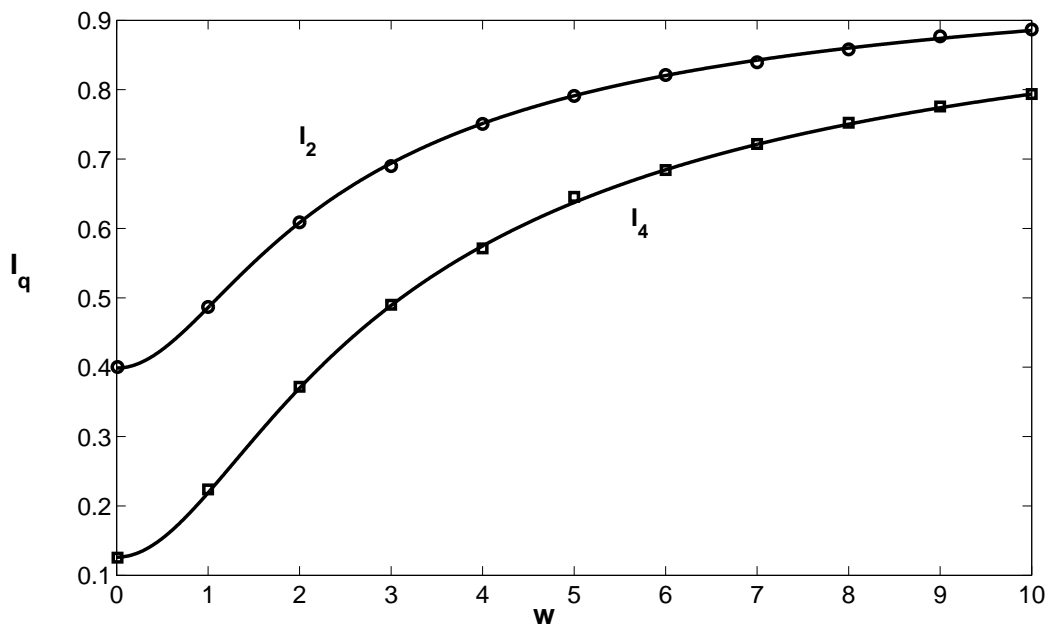}
\end{center}
\caption{Numerical (symbols) and analytical (solid lines) results for $I_2$ and $I_4$ as a function of $w$.}
\label{fig1}
\end{figure}

\begin{figure}[t]
\begin{center}
\includegraphics[clip=true,width=\columnwidth]{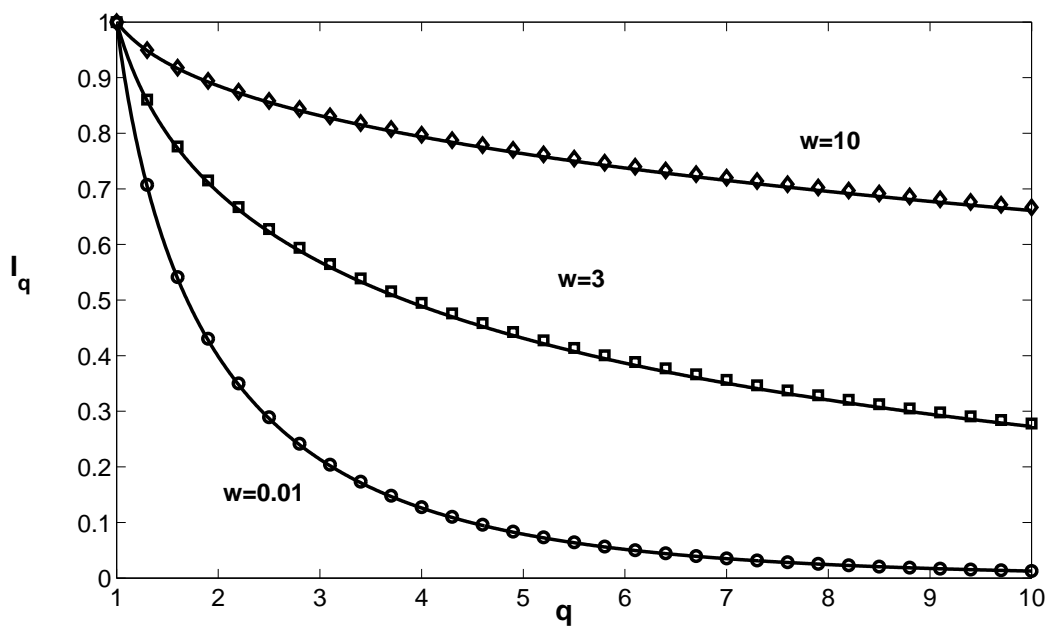}
\end{center}
\caption{Numerical (symbols) and analytical (solid lines) results for $I_q$ as a function of $q$ for $w=0.01,3$ and $10$.}
\label{fig2}
\end{figure}

In order to test the validity of Eq.(\ref{Iq_final}) and Eq.(\ref{Iq_all_q}) at different values of $q$ and  $w$ we compare them with the results of numerical simulations. The moments of the eigenvectors were calculated by direct diagonalization of $500\times 500$ random matrices defined by Eq.(\ref{ham}). The number of realizations was chosen in such a way, that $5000$ different eigenvectors with eigenvalues close to $E=0$ were generated for each value of $w$ and $q$. In Fig.~\ref{fig1} the results for $I_2$ and $I_4$ for various disorder strengths $w$ are compared with Eq.(\ref{Iq_final}). The numerical and the analytical results for $I_q$ as a function of $q$ are presented in Fig.~\ref{fig2} for $w=0.01,3$ and $10$. Here the analytical predictions were obtained from Eq.(\ref{Iq_all_q}). One can notice that the results of numerical simulations are in excellent agreement with the analytical predictions for all values of $w$ and $q$.

%**************************************************************************************

\section{Conclusions}\label{sec_concl}

We have presented a field-theoretical representation for the moments of the eigenstates, which involves a single real-valued field variable. The representation is exact and can be applied to a generic Anderson model with non-random hopping elements. The action of the constructed field theory is non-local, due to the presence of the functional determinant. In certain cases, the functional determinant can be calculated exactly. Generally, our representation can serve as a starting point for further 
approximate calculations, which might be efficient, for example, in the limit of strong or weak disorder. 

We would like to point out that the derived field theory shares some similarities with the partition function of the hyperbolic sigma model studied recently in Ref.\cite{DSZ10}. Indeed, if we restrict all $t_p$ in Eq.(\ref{gen_fun_final}) to be positive and replace them by $t_p=e^{\theta_p}$, then the action expressed in the new variables $\theta_p$ is similar to the action of the  hyperbolic sigma model in the horospherical coordinate system. Therefore, it might be possible to apply the methods from Ref.\cite{DSZ10} to prove rigorously the existence of the diffusive phase in the three-dimensional Anderson model.

We have applied the general formula for the  moments of the eigenstates to the simplex model and derived a very compact representation for the moments. Our analysis has shown that the moments remain finite in the thermodynamic limit implying that the eigenstates are localized at any strength of disorder. The analytical expression for the moments that we obtained agrees completely with the results of the numerical simulations.

The fact that the eigenstates are always localized might be in conflict with the expectation that the presence of the hopping term, which connects any two sites of the lattice, must lead to delocalization at least at weak disorder. In order to understand, why this is not the case, one can consider first the opposite limit of strong disorder. If disorder is sufficiently strong, then we expect that the eigenstates are localized. Treating the hopping term $T$ as a perturbation, we can notice that it is a rank one matrix, as it is given by $T=(1/N)\ket{s}\bra{s}$, where $\ket{s}$ and $\bra{s}$ are defined by Eq.(\ref{Dirac}). One can argue that the rank one perturbation remains always week independently of the strength of its matrix elements. Indeed, one can show that our approach can be used for any rank one hopping matrix and the results will be qualitatively the same. We would like to stress that the fact, that the hopping elements are non-random, is crucial for this conclusion.

Another way to explain the observed localization of the eigenstates is to notice that, in contrast to the standard Anderson model, the eigenstates of the simplex model are degenerate at $w=0$. The presence of disorder of the strength $w$ leads to the appearance of the energy band of the width of the order of $w$, as it follows from the expression for the density of states $\rho\sim 1/w$. Hence the bandwidth is always of the same order as the disorder strength, implying that disorder is actually strong at any value of $w$ \cite{Kravtsov}.

%*****************************************************************************************************************************
%*****************************************************************************************************************************
\appendix
\section{Integration over the phases}\label{app_phase}
In this Appendix we calculate the integral over the phases
\begin{eqnarray} \label{J-int}
\hspace*{-40pt}J&=&\int_0^{2\pi}\prod_{p\neq n}d\phir(p)\: d\phia(p)
%\exp&&\left[i \sum_{p \neq q}T_{pq}
e^{i \sum_{p,q}T_{pq}
\left(\cos(\phir(p)-\phir(q))g_{pq}^{R}
%-\right.\right.\nonumber\\
%&&\left.\left.
-\cos(\phia(p)-\phia(q))g_{pq}^{A}\right)
},\nonumber\\
%\right)\right]\\
\hspace*{-40pt}g_{pq}^{R}&=&\frac{\sqrt{s_ps_q}}{2\epsilon}
\left(1+\frac{\epsilon}{2}\left(\frac{v_p}{s_p}+\frac{v_q}{s_q}\right)\right),\;
g_{pq}^{A}=\frac{\sqrt{s_ps_q}}{2\epsilon}
\left(1-\frac{\epsilon}{2}\left(\frac{v_p}{s_p}+\frac{v_q}{s_q}\right)\right),
\end{eqnarray}
using the stationary phase approximation, which gives an exact result in the limit $\epsilon\to 0$. The stationary phase condition applied to the integral (\ref{J-int}) over $\phir(p)$ yields
\be\label{phi_system}
\sum_{q}T_{pq}\sin(\phir(p)-\phir(q))g_{pq}^{R}=0,
\en
which should be satisfied for all $p=1,\dots, N$ except for $p=n$. This system of equations has obvious solutions
\be\label{triv_sol}
\left[\begin{array}{c}\phir(p)-\phir(q)=0\\
\phir(p)-\phir(q)=\pi
\end{array}\right.,\quad {\rm if}\; T_{pq}\neq 0.
\en
Taking into account that $\phir(n)=0$, we conclude that each variable $\phir(p)$ has two stationary phase values $0$ and $\pi$, which it can take independently from a value of any other variable $\phir(q)$. It is convenient to introduce the new variables 
\be
\sigma_R(p)=\left\{\begin{array}{c}1,\quad \phir(p)=0\\
-1,\quad \phir(p)=\pi
\end{array}\right.,
\en
such that the $\phir$-dependent part of the action can be expanded around the stationary phase values as
\begin{eqnarray}\label{phi_expan}
\hspace*{-40pt}g_{pq}^{R}\cos(\phir(p)-\phir(q))&=&\sigma_R(p)\sigma_R(q)\sqrt{s_ps_q}\times\nonumber\\
\hspace*{-40pt}&&\left(\frac{1}{2\epsilon}-\frac{1}{4\epsilon}(\phir(p)-\phir(q))^2+
\frac{1}{4}\left(\frac{v_p}{s_p}+\frac{v_q}{s_q}\right)\right),
\end{eqnarray}
where we neglect all higher order terms in $(\phir(p)-\phir(q))$, as they give no contribution in the limit $\epsilon\to 0$.

Repeating the same steps for the integral over $\phia(p)$, we obtain
\begin{eqnarray}
\hspace*{-40pt}g_{pq}^{A}\cos(\phia(p)-\phia(q))&=&\sigma_A(p)\sigma_A(q)\sqrt{s_ps_q}\times\nonumber\\
\hspace*{-40pt}&&\left(\frac{1}{2\epsilon}-\frac{1}{4\epsilon}(\phia(p)-\phia(q))^2-
\frac{1}{4}\left(\frac{v_p}{s_p}+\frac{v_q}{s_q}\right)\right),
\end{eqnarray}
Comparing these two results, we notice that the singular $1/\epsilon$ contribution to the action is canceled out iff 
$\sigma_R(p)\sigma_R(q)=\sigma_A(p)\sigma_A(q)$. Taking into account that $\sigma_R(n)=\sigma_A(n)=1$ by definition, we conclude that $\sigma_R(p)=\sigma_A(p)\equiv\sigma_p$ for all $p$.

For a fixed configuration of $\{\sigma_p\}$, the integral over $\phir(p)$ is given now by the Gaussian integral in the leading order in $\epsilon$:
\begin{eqnarray}\label{phir_gauss}
\int_{-\infty}^{\infty}\prod_{p\neq n}d\phir(p)e^{-\frac{i}{4\epsilon}
\sum_{p,q}\sigma_p\sigma_q\sqrt{s_ps_q}(\phir(p)-\phir(q))^2}=
\frac{\left(e^{-i\frac{\pi}{4}}\sqrt{2\pi\epsilon}\right)^{N-1}}{\sqrt{\det Q}}.
\end{eqnarray}
The matrix $Q$ is a $(N-1)\times (N-1)$ matrix defined as
\be
\hspace*{-60pt}Q_{pq}=\sigma_p\sigma_q\sqrt{s_ps_q}
\left(-T_{pq}+\delta_{pq}\sum_r T_{pr}\sigma_p\sigma_r\sqrt{\frac{s_r}{s_p}}\right),
\quad p,q=1,\dots,N;\;p,q\neq n.
\en
One can notice that $Q$ can be factorized as
\be
Q=\Sigma B \Sigma,
\en
where $\Sigma_{pq}=\sigma_p\sqrt{s_p}\delta_{pq}$ and
\be
\hspace*{-40pt}B_{pq}=-T_{pq}+\delta_{pq}\sum_r T_{pr}\sigma_p\sigma_r\sqrt{\frac{s_r}{s_p}},
\quad p,q=1,\dots,N;\;p,q\neq n.
\en 
The factorization enables us to express the determinant of $Q$ as
\be
\det Q= \det B \prod_{p\neq n}s_p.
\en
The integral over $\phia(p)$ can be evaluated in a similar way:
\be\label{phia_gauss}
\int_{-\infty}^{\infty}\prod_{p\neq n}d\phia(p)e^{\frac{i}{4\epsilon}
\sum_{p,q}\sigma_p\sigma_q\sqrt{s_ps_q}(\phia(p)-\phia(q))^2}=
\frac{\left(e^{i\frac{\pi}{4}}\sqrt{2\pi\epsilon}\right)^{N-1}}{\sqrt{\det Q}}.
\en
Collecting the results from Eqs.(\ref{phi_expan}), (\ref{phir_gauss}) and (\ref{phia_gauss}) we find
\be
J=(2\pi\epsilon)^{(N-1)}\sum_{\{\sigma_p\}}e^{i\sum_{p,q}T_{pq}\sigma_p \sigma_q v_p\sqrt{\frac{s_q}{s_p}}}
\frac{1}{\det B}\:\prod_{p\neq n}\frac 1s_p.
\en
We assume that the system of equations (\ref{phi_system}) has only trivial solutions given by Eq.(\ref{triv_sol}), which is the case for a generic matrix $T_{pq}$. If for a specific choice of $T_{pq}$ a non-trivial solution of (\ref{phi_system}) does exist, its contribution to the integral should be also taken into account in a similar way. 
%*****************************************************************************************************

\section{Integration over the Grassmann variables}\label{app_grass}
The integral over the Grassmann variables (\ref{G_int}) can be factorized as follows:
\begin{eqnarray}\label{G_int_app}
G&=&\left(\int \prod_{p\neq n}[d \chi_p\chi_p^{\ast}]e^{i\sum_{p,q}C_{pq}
\chi_p^{\ast}\chi_p}\right)^2,\nonumber\\
C_{pq}&=&T_{pq}-\delta_{pq}\sum_{r}T_{pr}\sigma_p\sigma_r \sqrt{\frac{s_r}{s_p}}.
\end{eqnarray}
This integral is Gaussian, but the quadratic form  
$\sum_{p,q}C_{pq}\chi_p^{\ast}\chi_p$ contains variables $\chi_n$ and $\chi_n^{\ast}$, which are not integrated out. In order to eliminate these two variables from the integrand, one can perform the change of the variables
\be
\chi_p=\eta_p+D_p\chi_n,\;\chi_p^{\ast}=\eta_p^{\ast}+D_p\chi_n^{\ast},\quad D_p=\sigma_p\sigma_n\sqrt{\frac{s_p}{s_n}}.
\en
In terms of the new variables the quadratic form reads
\begin{eqnarray}\label{quadr_form}
\hspace*{-60pt}\sum_{p,q}C_{pq}\chi_p^{\ast}\chi_q&=&
\sum_{pq}^{\prime}C_{pq}\eta_p^{\ast}\eta_q+
\left(\sum_{p,q}^{\prime}C_{pq}D_q\eta_p^{\ast}+\sum_p^{\prime}C_{pn}\eta_p^{\ast}\right)\chi_n+\nonumber\\
&&\hspace*{-80pt}
\chi_n^{\ast}\left(\sum_{p,q}^{\prime}C_{pq}D_p\eta_q+\sum_p^{\prime}C_{pn}\eta_p\right)+
\left(\sum_{p,q}^{\prime}C_{pq}D_pD_q+2\sum_p^{\prime}C_{pn}D_p+C_{nn}\right)\eta_n^{\ast}\eta_n,
\end{eqnarray}
where the symbol $\sum^{\prime}$ means that the terms with $p=n$ and $q=n$ must be excluded from the sum. 

Now we show that all the terms, except for the first one, on the right-hand side of the equation above are equal to zero. Indeed, it follows from the definition of $C_{pq}$, that for $p\neq n$
\begin{eqnarray}
\hspace*{-40pt}\sum_{q}^{\prime}C_{pq}D_q+C_{pn}&=&
\sum_{q}^{\prime}\left(T_{pq}-\delta_{pq}\sum_r T_{pr}\sigma_p\sigma_r \sqrt{\frac{s_r}{s_p}}\right)
\sigma_q\sigma_n \sqrt{\frac{s_q}{s_n}}+T_{pn}=\nonumber\\
\hspace*{-40pt}&& \sum_{q}^{\prime}T_{pq}\sigma_q\sigma_n \sqrt{\frac{s_q}{s_n}}-
\sum_r T_{pr}\sigma_r\sigma_n \sqrt{\frac{s_r}{s_n}}+T_{pn}=0.
\end{eqnarray}
Hence the second and the third terms in Eq.(\ref{quadr_form}) vanish. Similarly, for the last term in Eq.(\ref{quadr_form})
we obtain\begin{eqnarray}
\hspace*{-40pt}\sum_{p,q}^{\prime}C_{pq}D_pD_q&=&
\sum_{q}^{\prime}C_{pq}\left(T_{pq}-\delta_{pq}\sum_r T_{pr}\sigma_p\sigma_r \sqrt{\frac{s_r}{s_p}}\right)
\sigma_p\sigma_n \sqrt{\frac{s_p}{s_n}}\sigma_q\sigma_n \sqrt{\frac{s_q}{s_n}}=\nonumber\\
&& \hspace*{-40pt}\sum_{p,q}^{\prime}T_{pq}\sigma_p\sigma_q \frac{\sqrt{{s_p}{s_q}}}{s_n}-
\sum_p^{\prime}\sum_r T_{pr}\sigma_p\sigma_r \frac{\sqrt{{s_p}{s_q}}}{s_n}=
-\sum_{p}^{\prime}T_{pn}\sigma_p\sigma_n \sqrt{\frac{s_p}{s_n}}.
\end{eqnarray}
On the other hand, the sum of the remaining two terms is equal to
\begin{eqnarray}
\hspace*{-40pt}2\sum_{p}^{\prime}C_{pn}D_p+C_{nn}&=&
2\sum_{p}^{\prime}T_{pn}\sigma_p\sigma_n \sqrt{\frac{s_p}{s_n}}+T_{nn}-
\sum_{r}T_{rn}\sigma_r\sigma_n \sqrt{\frac{s_r}{s_n}}
=\nonumber\\
&&\sum_{p}^{\prime}T_{pn}\sigma_p\sigma_n \sqrt{\frac{s_p}{s_n}},
\end{eqnarray}
so that we conclude that the last term in Eq.(\ref{quadr_form}) equals to zero. Thus, we find from Eq.(\ref{quadr_form})
\be
\sum_{p,q}C_{pq}\chi_p^{\ast}\chi_p=
\sum_{p,q}^{\prime}C_{pq}\eta_p^{\ast}\eta_q=-\sum_{p,q}^{\prime}B_{pq}\eta_p^{\ast}\eta_q,
\en
where $B_{pq}$ is defined by Eq.(\ref{B_matr}). The integral in (\ref{G_int_app}) is now the standard Gaussian one, and we immediately obtain
the result
\be
G=(-1)^{N-1}(\det B)^2.
\en
%*****************************************************************************************************

\section{Calculation of the functional determinant}\label{app_det}
The matrix $B_{pq}$ defined by Eq.(\ref{B_matr_t}) is the $(N-1)\times (N-1)$ matrix, which can be obtained 
from the  $N\times N$ matrix $\hb$
\be
\hspace*{0pt}\hb_{pq}=
-T_{pq}+\delta_{pq}\sum_r T_{pr}\frac{t_r}{t_p},
\quad p,q=1,\dots,N,
\en 
by eliminating all the elements in $n$th row and in the $n$th column. The matrix $\hb$ is singular, as the vector
\be\label{zero_mode}
z=\left(\sum_{p=1}^{N}t_p^2\right)^{-1/2}(t_1,t_2,\dots,t_N)^{T} 
\en
is the normalized eigenvector of $\hb$ corresponding to the zero eigenvalue. 

Using the first-order  perturbation theory for the eigenvalues, it is easy to show that
\be
\det \hb^{pq}=(-1)^{p+q} z_p^{\ast}z_q\: \overline{\det}\hb,
\en  
where $\hb^{pq}$ is an $(N-1)\times (N-1)$ matrix, obtained from $\hb$ by eliminating all the elements in $p$th row and in the $q$th column and $\overline{\det}\hb$ is the product of all non-zero eigenvalues of $\hb$. This formula is valid for an arbitrary singular matrix $\hb$, which has the non-degenerate zero eigenvalue with an eigenvector $z$. In  particular, it implies that 
\be\label{detB}
\det B= \abs{z_n}^2\odet\hb
\en
On the other hand, the product of all non-zero eigenvalues of $\hb$ can be calculated as
\be\label{odet}
\odet \hb = \lim_{\epsilon\to 0}\frac{1}{\epsilon} \det (\hb+\epsilon I),
\en
where $I$ is the $N\times N$ identity matrix. This result simply follows from the fact that adding of $\epsilon I$  to $\hb$ shifts all the eigenvalues of $\hb$, including the zero one, by $\epsilon$. 

For the simplex model, the matrix $\hb$ reads
\be
\hb_{pq}=\left\{\begin{array}{c}\frac{1}{N t_p}\sum_{r=1}^N t_r-\frac{1}{N},\quad p=q\\
-\frac{1}{N},\quad p\neq q
\end{array}\right.,
\en
so that one can represent it using the Dirac notation as follows
\begin{eqnarray}\label{Dirac}
\hspace*{-60pt}\hb&=&\frac{1}{N}(-\ket{s}\bra{s}+\alpha D),\\
\hspace*{-60pt}\bra{s}&=&(1,1,\dots,1),\;\ket{s}=(1,1,\dots,1)^{T},\; \alpha=\sum_{r=1}^N t_r,
\; D=\diag (t_1^{-1},t_2^{-1},\dots,t_N^{-1}).
\end{eqnarray}
According to Eq.(\ref{odet}) we need to compute 
\begin{eqnarray}
\label{prod_det}\det{(\hb+\epsilon I)}=\frac{1}{N^N}\det (\te I+\alpha D)\: \det (I-G_0 \ket{s}\bra{s}),&&\\
G_0=(\te I+\alpha D)^{-1},&&
\end{eqnarray}
where $\te$ stands for $N\epsilon$. The first determinant on the right-hand side of Eq.(\ref{prod_det}) can be found easily, as the matrix 
$\te I+\alpha D$ is diagonal:
\be\label{diag_det}
\det (\te I+\alpha D)=\prod_{p=1}^N\left(\frac{\alpha}{t_p}+\te\right).
\en 
In order to calculate the second determinant on the right-hand side of Eq.(\ref{prod_det}) we notice that
\begin{eqnarray}\label{ln_det}
\hspace*{-40pt}\ln \det (I-G_0 \ket{s}\bra{s})=\tr \ln (I-G_0 \ket{s}\bra{s})=-\sum_{n\ge 1}\frac{1}{n}\tr
\left(G_0\ket{s}\bra{s}\right)^n=&&\nonumber\\
\hspace*{-40pt}-\sum_{n\ge 1}\frac{1}{n}\left(\bra{s}G_0\ket{s}\right)^n=\ln(1-\bra{s}G_0\ket{s})=
\ln\left(1-\sum_{p=1}^N\left(\frac{\alpha}{t_p}+\te\right)^{-1}\right).&&
\end{eqnarray}
From Eqs.(\ref{prod_det}), (\ref{diag_det}) and (\ref{ln_det}) we find that
\be
\det{(\hb+\epsilon I)}=\frac{1}{N^N}
\left(1-\sum_{p=1}^N\left(\frac{\alpha}{t_p}+\te\right)^{-1}\right)
\prod_{p=1}^N\left(\frac{\alpha}{t_p}+\te\right).
\en
Then it follows from Eq.(\ref{odet}) that
\begin{eqnarray}
\odet \hb&=&\lim_{\epsilon\to 0}\frac{1}{\epsilon} 
\frac{1}{N^N}\left(1-\sum_{p=1}^N\left(\frac{\alpha}{t_p}+\te\right)^{-1}\right)
\prod_{p=1}^N\left(\frac{\alpha}{t_p}+\te\right)=\nonumber\\
&&\frac{1}{N^{N-1}}\sum_{r=1}^N t_r^2 \left(\sum_{r=1}^N t_r \right)^{N-2}\prod_{p=1}^N\frac{1}{t_p}.
\end{eqnarray}
Substituting this result into Eq.(\ref{detB}) and taking into account Eq.(\ref{zero_mode}) we obtain
finally the expression for the determinant of the matrix $B$:
\be\label{det_simplex}
\det B= \frac{t_n^2}{N^{N-1}} \left(\sum_{r=1}^N t_r \right)^{N-2}\prod_{p=1}^N\frac{1}{t_p}.
\en
%*****************************************************************************************************

\section{Calculation of the density of states}\label{app_dos}

The averaged density of states is defined as
\be
\rho (E)=\frac{1}{N}\sum_{p=1}^N \avg{\delta\left(E-E_p\right)}=\frac{1}{\pi}{\rm Im} \avg{G(E-i0)},
\en 
where $i0$ is an infinitesimal imaginary shift in energy and the Green's function $G$ is given by
\be
G({\cal E})=\frac1N\tr \frac{1}{{\cal E}-H}.
\en
For the simplex model, the Green's function reads
\be\label{GG}
G({\cal E})=\tr(G_0^{-1}-\ket{s}\bra{s})^{-1},
\en
where $(G_0^{-1})_{pq}=\delta_{pq}({\cal E}-H_{pp})N$ and $\ket{s}$ and $\bra{s}$ are defined by Eq.(\ref{Dirac}). 
Since $G_0$ is diagonal and $\ket{s}\bra{s}$ is an operator of rank one, the inverse operator in Eq.(\ref{GG}) can be calculated explicitly:
\begin{eqnarray}
&&G({\cal E})= \tr (I-G_0 \ket{s}\bra{s})G_0=\tr \sum_{k\ge 0} (G_0 \ket{s}\bra{s})^kG_0=\nonumber\\
&& \tr\left(G_0+\sum_{k\ge 1}(\bra{s} G_0 \ket{s})^{k-1}G_0\ket{s}\bra{s}G_0\right)=
\tr G_0+\frac{\bra{s} G_0^2 \ket{s}}{1-\bra{s} G_0 \ket{s}}
\end{eqnarray}
The first term here corresponds to the diagonal part of the Hamiltonian and its contribution to the density of states is equal to $P(E)$, where $P(H_{pp})$ is the probability distribution of the diagonal elements. One can show that the contribution of the second term vanishes in the limit $N\to \infty$ due to the presence of $G_0^2$ in the numerator. As a result, we obtain
\be
\rho (0) =P(0)=\frac{1}{\sqrt{2\pi} w}.
\en 

%*****************************************************************************************************

\end{document}